\documentstyle[12pt]{article}

\begin{document}
\begin{titlepage}

\hfill
TSU/QFTD-36/92
\vspace{1cm}

\hfill
28th December 1992
\vspace{1cm}
\begin{center}
{\bf\LARGE Higher spins dynamics \\in the closed string theory\\}
\end{center}
\vspace{3 cm}
\noindent
{\large I.L. BUCHBINDER\\
{\it Department of Theoretical Physics,\\
Tomsk State Pedagogical Institute, Tomsk 634041, Russia\\}}
\vspace{1cm}

\noindent
{\large E.S. FRADKIN\\
{\it Department of Theoretical Physics,\\
 P.N. Lebedev Physical Institute, Moscow 117924, Russia\\}}
\vspace{1cm}

\noindent
{\large S.L. LYAKHOVICH and V.D. PERSHIN \\
{\it Department of Quantum Field Theory, \\
Tomsk State University, Tomsk 634050, Russia}}

\vspace{3.0cm}

\begin{abstract}
The general $\sigma$-model-type string action including both massless
and massive higher spins background fields is suggested. Field equations
for background fields are followed from the requirement of quantum Weyl
invariance. It is shown that renormalization of the theory
can be produced level by level. The detailed consideration of
background fields structure and corresponding fields equations is given
for the first massive level of the closed bosonic string.
\end{abstract}
\vfill\null
\end{titlepage}

\newpage

{\bf 1. Introduction.}
The $\sigma$-model approach \cite{1,2} still remains a basic method of
description of string modes interactions allowing to construct effective
action for string excitations \cite{2}. This approach is based on insertion
of the interaction with background fields into string action and on the
principle of Weyl invariance of corresponding quantum theory. The
programme of investigation of different aspects of string theory in
framework of such approach has been proposed in \cite{1}-\cite{3}
and considered in
numerical publications (for review see \cite{4,5} ) but for massless
background only. As a result the above approach has led to the famous
equations for background fields.

This paper is devoted to a generalization of Fradkin-Tseytlin effective
action approach \cite{2} to describe not only massless but also massive
higher spin modes dynamics. By the present time different approaches to
derivation linear field equations for massive string excitations have been
developed \cite{6}-\cite{11}\footnote{We do not discuss here the interaction
of the tachion mode (see for example \cite{9,13})}
We hope that our approach allows to obtain interacting equations for
massive higher spins fields\footnote{The problem of higher spin
interactions has been investigated recently in \cite{12} on the base of new
supersymmetry out of string theory framework.}.

The standart $\sigma$-model lagrangian \cite{1,2} is constructed with the
help of dimension two operators $\partial_a x^\mu \partial_b x^\nu$ and
$R^{(2)}$, where $x^\mu(\sigma)$ are  string coordinates and $R^{(2)}$ is
world sheet curvature. The corresponding irreducible background fields
$G_{\mu\nu}(x)$, $A_{\mu\nu}(x)$ and $\phi(x)$ are identified with massless
states in the closed string spectrum. On the face of it the inclusion of
massive mode background fields is a trivial enough problem. We should simply
add to standart $\sigma$-model lagrangian terms containing
derivatives of fourth, sixth, eighth and other orders with respect to world
sheet coordinates $\sigma^a$. This procedure has been discussed for many times
(see for examples \cite{2,9,14,15}).

It is convenient to rescale string coordinates $x^\mu
\rightarrow \sqrt{\alpha'} x^\mu$. Then  taking
into account contributions from all string modes, both massless and
massive, the complete action should have the form
\begin{eqnarray}
S = \int \! d^2 \sigma  \; \sum_{n=0}^\infty (\alpha')^n
    \sum_{i_n = 1}^{N_n}
    {\cal O}_{i_n}^{(n)} \bigl( \sigma, \partial x (\sigma) \bigr) \;
    B_{i_n}^{(n)}  \bigl( x (\sigma) \bigr)
\end{eqnarray}
Here $B^{(n)}_{i_n}$ are  background fields corresponding to given massive
level $n$ and ${\cal O}^{(n)}_{i_n}$ are some composite operators constructed
from $\sqrt{g}$, $g^{ab}$, $\varepsilon^{ab}$, $\partial x$, $R^{(2)}$,
$\partial \partial x$, $\partial R^{(2)}$, \ldots . At given $n$ the dimension
of
${\cal O}^{(n)}_{i_n}$ in derivatives with respect to $\sigma^a$ is equal to
$2n+2$. $N_n$ is the number of irreducible operators ${\cal O}^{(n)}_{i_n}$ at
given $n$. All background fields $B^{(n)}_{i_n}$  as well as
the string coordinates $x^\mu (\sigma)$ are dimensionless.
Now the power of $\alpha'$ is
connected with the number of given massive level. Hence by following the power
of $\alpha'$ we follow the corresponding massive level contribution.

But if we try to consider the theory with the action (1) seriously we will
face immediately with the following problems.

\begin{itemize}
\item{a.} Is it really possible to write out all irreducible operators
${\cal O}^{(n)}_{i_n}$ and corresponding background fields $B^{(n)}_{i_n}$ for
each given massive level?
\item{b.} May be we will not have enough background fields to find
correspondence with the string spectrum or, on the contrary, will have too
many of them.
\item{c.} It is evident that the theory under consideration will be
nonrenormalizable if we take into account  any finite number of massive
levels in the action (1).
One can expect that the theory will be renormalizable only in generalized
sense when  infinite number of terms is included into the action together
with infinite number of all admissible counterterms. As the action (1)
includes all permissible operators constructed of $\partial_a$, $x^\mu$ and
$g_{ab}$ with all possible background fields we believe that the theory will
be multiplicatively renormalizable because all suitable counterterms
should have structure analogous to the structure of action (1). It
means that we have a possibility to construct renormalized
trace of  energy-momentum tensor and to get  equations of motion for
background fields $B^{(n)}_{i_n}$ by imposing the condition of Weyl
invariance.
Unfortunately, this programme has a weak point. It is not a simple thing to
calculate the quantum trace of energy-momentum tensor in explicit form taking
into account contributions of all massive levels. It seems to be more
suitable to consider massive levels in turn step by step, starting with the
massless level and then adding the first one, the second and so on. We will
show that such point of view really can be realized.
\item{d.} If we demand the theory under consideration
to have quantum Weyl invariance
then we can expect some equations for  background fields.
Can we be sure that these equations correspond to the known equations
defining each given string massive level? Obviously, to answer this question
it is sufficient to deduce linear equations for background fields.
\item{e.} One can expect that the requirement of Weyl invariance will
lead to equations of motion for bosonic fields of lower and higher spins
interacting with each other. It is generally accepted that these equations are
inconsistent. Hence we can face with the inconsistency problem in the string
theory as well.
\end{itemize}

So, the attempt to add to standart $\sigma$-model action terms
corresponding to massive string excitations immediately creates the essential
problems. The main aim of this paper is to try to consider some of
them in details.

 \vspace{0.5cm}
{\bf 2. Structure of One-loop Renormalization.}
The main result we are going to discuss here is formulated as follows. Let us
consider the action (1) where all massive level contributions are taken into
account and calculate one-loop counterterms. Then to find  counterterms
giving contributions to renormalization of definite massive level it is
sufficient to take into account only this given level and all lower massive
levels. The background fields of higher massive levels can not give
contribution to renormalization of the given level background fields.

Let us write one-loop effective action:
$$
\Gamma^{(1)} = S + {1\over{2}} \mbox{Tr} \ln {\cal H} ,
$$
where $S$ is the action (1) and the operator ${\cal H}_{\mu\nu}$ is
\begin{eqnarray}
{\cal H}_{\mu\nu} = {\delta^2 S \over {\delta x^\mu \delta x^\nu} }.
\end{eqnarray}
It is evident that this operator has the form
\begin{eqnarray}
{\cal H} \; \sim \; - D^2 + \sum_{k=0}^\infty P^{a_1 \ldots a_{2k}}
                             D_{a_1} \ldots D_{a_{2k}},
\end{eqnarray}
where
$$
D_a \partial_b x^\mu =        \partial_a   \partial_b x^\mu -
                       \Gamma^c_{ab} (g)   \partial_c x^\mu +
{\bf \Gamma}^\mu_{\lambda\rho} (G) \partial_a x^\lambda \partial_b x^\rho ,
$$
$ \Gamma (g) $ and $ {\bf \Gamma} (G) $ are connections constructed of the
world sheet metric $g_{ab}$ and background metric $G_{\mu\nu}$ respectively,
$D^2 = g^{ab} D_a D_b$. The coefficients $P^{a_1 \ldots a_{2k}}$ depend on
background fields and can be written as
\begin{eqnarray}
P = \sum_{n=0}^\infty (\alpha')^n P^{(n)}.
\end{eqnarray}
Here $P^{(n)}$ depend only on background fields of the given massive
level with the number $n$. Then
\begin{eqnarray}
{\rm Tr} \ln {\cal H} = {\rm Tr} \ln \left( - D^2 \right)
          - \sum_{l=1}^\infty {1 \over l} {\rm Tr} \left(
\sum_{k=0}^\infty P^{a_1 \ldots a_{2k}} D_{a_1} \ldots D_{a_{2k}}
             {1 \over D^2} \right)^l
\end{eqnarray}
Remind that all background fields as well as coordinates $x^\mu$ are
dimensionless. The only dimensional parameters of the theory are $\alpha'$ and
$\sigma^a$.

The action corresponding to $n$-th massive level is
\begin{eqnarray}
  (\alpha')^n \int \! d^2 \sigma
    \sum_{i_n = 1}^{N_n}  {\cal O}_{i_n}^{(n)}   B_{i_n}^{(n)}
\end{eqnarray}
A suitable counterterm should have the following form
\begin{eqnarray}
  (\alpha')^n \int \! d^2 \sigma
\sum_{i_n = 1}^{N_n}  {\cal O}_{i_n}^{(n)}  {\cal T}_{i_n}^{(n)} (B),
\end{eqnarray}
where ${\cal T}_{i_n}^{(n)} (B)$ are some dimensionless functions of background
fields. Taking into account the eqs.(4,5) we see that to create
$(\alpha')^n$ in the counterterm (7) one can use in action (1)
only $n$-level and all other levels with numbers less than $n$.

Thus, to renormalize background fields of massless level we can use only
 background fields of this level. To renormalize  background fields of
the first level we can use fields of this level and of the massless one and
so on. Although the theory is renormalizable only if infinite number of
counterterms is taken into account, we can use these counterterms step by
step, level by level. Renormalization of any given level background fields
demands only the finite number of counterterms. Remind that we have made
the above
conclusion in the one-loop approximation. However we believe that this result
should valid in higher loops as well.

Thus we have a possibility to consider $\sigma$-model type action containing
contributions only of finite number of massive string levels. For example the
action in the theory of interacting massless and first massive level
excitations should have the form
\begin{eqnarray}
\tilde S = \int \! d^2 \sigma \left(
    \sum_{i_0 = 1}^{N_0}  {\cal O}_{i_0}^{(0)}   B_{i_0}^{(0)}
+ \alpha' \sum_{i_1 = 1}^{N_1}  {\cal O}_{i_1}^{(1)}   B_{i_1}^{(1)} \right)
\end{eqnarray}
The rest of the paper will be devoted to consideration of this case only.

\vspace{0.5cm}

{\bf 3. Background Fields of First Massive Level.}
Our aim here is to write out all irreducible operators ${\cal O}_{i_1}^{(1)}$
and to introduce all corresponding background fields $B_{i_1}^{(1)}$. The
operators ${\cal O}_{i_1}^{(1)}$ can be constructed of the objects
$\partial_a x^\mu$,
$D_a \partial_b x^\mu$, $D_a D_b \partial_c x^\mu$,
$D_a D_b D_c \partial_d x^\mu$, $R^{(2)}$, $\partial_a R^{(2)}$,
$D_a \partial_b R^{(2)} $ employing the metric
$g^{ab}$ and antisymmetric tensor $\varepsilon^{ab}$ and taking into
account that the dimension of the ${\cal O}_{i_1}^{(1)}$
should be equal to four. As a result we obtain:
\begin{eqnarray}
\sum_{i_1 = 1}^{N_1}  {\cal O}_{i_1}^{(1)}   B_{i_1}^{(1)} &=&
                     \sqrt{g} \left\{
g^{ab} g^{cd} \partial_a x^\mu \partial_b x^\nu \partial_c x^\lambda
\partial_d x^\kappa                      F^{(1)}_{\mu\nu\lambda\kappa} (x) +
\right.
\nonumber\\ & &{}+
g^{ab} \varepsilon^{cd} \partial_a x^\mu \partial_b x^\nu \partial_c x^\lambda
\partial_d x^\kappa                      F^{(2)}_{\mu\nu\lambda\kappa} (x) +
\nonumber\\ & &{}+
g^{ab} D^2 x^\mu \partial_a x^\nu \partial_b x^\lambda
                                               T^{(1)}_{\mu\nu\lambda} (x) +
g^{ac} g^{bd} D_a \partial_b x^\mu \partial_c x^\nu \partial_d x^\lambda
                                               T^{(2)}_{\mu\nu\lambda} (x) +
\nonumber\\ & &{}+
g^{ac} \varepsilon^{bd} D_a \partial_b x^\mu \partial_c x^\nu
\partial_d x^\lambda                           T^{(3)}_{\mu\nu\lambda} (x) +
\nonumber\\ & &{}+
D^2 x^\mu D^2 x^\nu                                   M^{(1)}_{\mu\nu} (x) +
g^{ac} g^{bd} D_a \partial_b x^\mu D_c \partial_d x^\nu
                                                      M^{(2)}_{\mu\nu} (x) +
\nonumber\\ & &{}+
g^{ac} \varepsilon^{bd} D_a \partial_b x^\mu D_c \partial_d x^\nu
                                                      M^{(3)}_{\mu\nu} (x) +
g^{ab} D_a D^2 x^\mu \partial_b x^\nu                 N^{(1)}_{\mu\nu} (x) +
\nonumber\\ & &{}+
\varepsilon^{ab} D_a D^2 x^\mu \partial_b x^\nu       N^{(2)}_{\mu\nu} (x) +
D^2 D^2 x^\mu                                                    V_\mu (x) +
\nonumber\\ & &{}+
R^{(2)} g^{ab} \partial_a x^\mu \partial_b x^\nu      W^{(1)}_{\mu\nu} (x) +
R^{(2)} \varepsilon^{ab} \partial_a x^\mu \partial_b x^\nu
                                                      W^{(2)}_{\mu\nu} (x) +
\nonumber\\ & &{}+
g^{ab} \partial_a x^\mu \partial_b R^{(2)}                 Y^{(1)}_\mu (x) +
\varepsilon^{ab} \partial_a x^\mu \partial_b R^{(2)}       Y^{(2)}_\mu (x) +
\nonumber\\ & &\left. {}+
R^{(2)} D^2 x^\mu                                                U_\mu (x) +
D^2 R^{(2)}                                                          Q (x) +
R^{(2)} R^{(2)}                                                      C (x)
                          \right\}
\end{eqnarray}
Here
$F^{(1)}_{\mu\nu\lambda\kappa} (x)$, $F^{(2)}_{\mu\nu\lambda\kappa} (x)$,
$T^{(1)}_{\mu\nu\lambda} (x)$, $T^{(2)}_{\mu\nu\lambda} (x)$,
$T^{(3)}_{\mu\nu\lambda} (x)$, $M^{(1)}_{\mu\nu} (x)$, $M^{(2)}_{\mu\nu} (x)$,
$M^{(3)}_{\mu\nu} (x)$, $N^{(1)}_{\mu\nu} (x)$, $N^{(2)}_{\mu\nu} (x)$,
$V_\mu (x)$, $W^{(1)}_{\mu\nu} (x)$, $W^{(2)}_{\mu\nu} (x)$,
$Y^{(1)}_\mu (x)$, $Y^{(2)}_\mu (x)$, $U_\mu (x)$, $Q (x)$, $C (x)$
are the background fields. Thus the eq.(9) contains the sum of eighteen
irreducible terms. Any other expression of dimension four can be written as a
linear combination of the above eighteen terms up to a total divergence.

Let us notice that we can add to the lagrangian (9) any total divergence and
so change the background fields. The complete list of divergences of
dimension four is
\begin{eqnarray}
& &D_a \left[ g^{ab} g^{cd} \partial_b x^\mu \partial_c x^\nu
\partial_d x^\lambda               \Lambda^{(1)}_{\mu\nu\lambda} (x) \right.
+ \varepsilon^{ab} g^{cd} \partial_b x^\mu \partial_c x^\nu
\partial_d x^\lambda               \Lambda^{(2)}_{\mu\nu\lambda} (x)
\nonumber\\ & &{}+
g^{ab} D^2 x^\mu \partial_b x^\nu
                                          \Lambda^{(3)}_{\mu\nu} (x)
+ g^{ab} g^{cd} D_b \partial_c x^{(\mu} \partial_d x^{\nu)}
                                          \Lambda^{(4)}_{\mu\nu} (x)
\nonumber\\ & &{}+
\varepsilon^{ab} D^2 x^\mu \partial_b x^\nu
                                          \Lambda^{(5)}_{\mu\nu} (x)
+ \varepsilon^{ab} g^{cd} D_b \partial_c x^{[\mu} \partial_d x^{\nu]}
                                          \Lambda^{(6)}_{\mu\nu} (x)
\nonumber\\ & &{}+
g^{ab} D_b D^2 x^\mu                           \Lambda^{(7)}_\mu (x)
+ g^{ab} \partial_b x^\mu R^{(2)}                \Lambda^{(8)}_\mu (x)
\nonumber\\ & &{}+\left.
\varepsilon^{ab} \partial_b x^\mu R^{(2)}
                                               \Lambda^{(9)}_\mu (x)
+ g^{ab} \partial_b R^{(2)}                     \Lambda^{(10)}_\mu (x)
                         \right]
\end{eqnarray}
It means there should be symmetry transformations of the background fields.
For example, the invariance of the action under adding the divergence
$D_a \left( g^{ab} g^{cd} \partial_b x^\mu \partial_c x^\nu
\partial_d x^\lambda \Lambda^{(1)}_{\mu\nu\lambda} (x) \right)$
to the lagrangian (9) leads to the following symmetry transformations:
\begin{eqnarray}
& & \delta F^{(1)}_{\mu\nu\lambda\kappa}=
        {1\over 2} \partial_{(\mu} \Lambda^{(1)}_{\nu)(\lambda\kappa)} +
        {1\over 2} \partial_{(\lambda} \Lambda^{(1)}_{\kappa)(\mu\nu)}
\nonumber\\ & &
\delta T^{(1)}_{\mu\nu\lambda} =
                                  \Lambda^{(1)}_{\mu(\nu\lambda)}
\nonumber\\ & &
\delta T^{(2)}_{\mu\nu\lambda} =
\Lambda^{(1)}_{\nu(\lambda\mu)} + \Lambda^{(1)}_{\lambda(\nu\mu)}
\end{eqnarray}
The analogous transformations can be written out for all other background
fields.

Let us notice that these transformations are not gauge ones since they do not
concern the dynamical variables $x^\mu$. However one can expect that the
requirement of the quantum Weyl invariance will lead to the equations of
motion for the background fields. Then the above transformations should be
the gauge transformations of those equations.

It is easy to see that these  transformations have the following general
structure
\begin{eqnarray}
          E \rightarrow E + \partial \Lambda,   \qquad
          S \rightarrow S + \Lambda
\end{eqnarray}
All background fields are divided into two classes -  ones of
$E$-type and  ones of $S$-type. The structure of transformations (12)
shows that  $S$-type fields are  Stueckelberg ones. Their rule is to
provide the gauge invariance in massive theory. Taking into account the
invariance under the above transformations we can fix the guage by setting
all the $S$-type fields equal to zero. The rest background fields are
$F^{(1)}_{\mu\nu\lambda\kappa} (x)$, $F^{(2)}_{\mu\nu\lambda\kappa} (x)$,
$W^{(1)}_{\mu\nu} (x)$, $W^{(2)}_{\mu\nu} (x)$, $C (x)$,
$T^{(1)}_{\mu\nu\lambda} (x)$,
$T^{(3)}_{\mu[\nu\lambda]} (x)$, $M^{(1)}_{\mu\nu} (x)$,
$U_\mu (x)$.

Our aim is to construct the quantum trace of energy-momentum tensor and then
to impose the requirement of the Weyl invariance. Let us consider the
contribution of the first massive level to the operator of trace. It presents
the sum of eighteen terms in accordance with the number of irreducible
four-dimensional operators. The coefficients at these operators are
combinations of the background fields and their derivatives. Unfortunately
the final result is complicated enough and we will enumerate only the
basic properties.
\begin{itemize}
\item{a.} Trace of the energy-momentum tensor is invariant under the above
symmetry transformations.

\item{b.} Trace of the energy-momentum tensor is equal to zero if and only if
all  massive background fields are equal to zero.

\item{c.} Background fields which are the coefficients at the $D^2 x^\mu$ in
the action (9) are present in the trace only as the coefficients at
$D^2 x^\mu$.
\end{itemize}

Let us consider the terms proportional to $D^2 x^\mu$ in the action (9). The
classical equations of motion are
$$
{\delta S \over \delta x^\mu} \sim G_{\mu\nu} D^2 x^\mu +
{\rm terms \; depending \; on \; background \; fields}
$$
Therefore all the terms $\sim D^2 x^\mu$ can be expressed through the
classical equations of motion with redefinition of some background fields.
But we know that the operator of classical equations of motion does not
contribute to quantum average values. So we can omit all the terms
$\sim D^2 x^\mu$ in the trace and hence in the action (9). The presence or
the absence of such terms does not influent on the quantum average value of
the trace operator.

Thus, setting all the $S$-type background fields equal to zero and omitting
terms $\sim D^2 x^\mu$ we will have the following list of essential
background fields:
$F^{(1)}_{\mu\nu\lambda\kappa} (x)$, $F^{(2)}_{\mu\nu\lambda\kappa} (x)$,
$W^{(1)}_{\mu\nu} (x)$, $W^{(2)}_{\mu\nu} (x)$, $C (x)$.

\vspace{0.5cm}

{\bf 4. Renormalization of Background Fields and Composite Operators in
Linear Approximation.}
As an illustration of the approach under consideration we will construct the
renormalized operator of energy-momentum tensor trace on the flat massless
background ($G_{\mu\nu} = \eta_{\mu\nu}$, $A_{\mu\nu} = 0$, $\phi = 0$)
taking into account only terms linear in massive background fields. We will
use the dimensional regularization and will start investigation from the
renormalization of the background fields.

In the one-loop approximation the divergent part of effective action can be
found from the following relation
\begin{eqnarray}
\left. \left(   {\rm Tr} \ln {\cal H}    \right) \right|_{div}
           &=& \left. {\rm Tr} \ln \left( - D^2 \right) \right|_{div}
  - {\rm Tr} \left(
       P^{a_1 a_2 a_3 a_4} D_{a_1} D_{a_2} D_{a_3} D_{a_4} {1\over{D^2}}
\right.
\nonumber \\& &
\left. \left.
 {}+   P^{a_1 a_2 a_3}  D_{a_1} D_{a_2} D_{a_3}     {1\over{D^2}} +
       P^{a_1 a_2}      D_{a_1} D_{a_2}         {1\over{D^2}} +
       P^{a_1}          D_{a_1}             {1\over{D^2}} +
       P                        {1\over{D^2}}
                   \right) \right|_{div}
\end{eqnarray}
The coefficients $P^{a_1 a_2 a_3 a_4}$, \ldots, $P^{a_1}$, $P$ are defined by
the eqs(3,4). The denotion $\left. (\ldots) \right|_{div}$ means that this
terms contain only the pole of the type $1 \over \epsilon$ where $\epsilon$
is a parameter of dimensional regularization. To extract the pole terms we
apply a two-dimensional analouge of the generalized Schwinger-DeWitt
technique \cite{16}.

The straightforward calculations lead to
\begin{eqnarray}
& &\left.{1 \over D^2} \right|_{div} =
               { {\mu^\epsilon} \over {2\pi\epsilon} } ,\qquad
\left.      D_{a_1} { 1 \over D^2} \right|_{div} =
\left. D_{a_1} D_{a_2} {1 \over D^2} \right|_{div} =
                                                   \ldots = 0
\end{eqnarray}
Omitting all calculational details we write out the final result for one-loop
divergences in the linear approximation:
\begin{eqnarray}
                 \left( \left.
     {\rm Tr} \ln {{\delta^2 S} \over {\delta x^\mu \delta x^\nu}}
                 \right) \right|_{div} =
    {\mu^\epsilon \over 2\pi\epsilon}  \int \! d^2 \sigma \sqrt{g}
                 \left(
  {d \over 6 } R^{(2)} - P^{\mu}{}_{\mu}
                 \right),
\end{eqnarray}
where $d$ is a tangent space dimension and $P^{\mu\nu}$ is the coefficient
$P_{\mu\nu}{}^{a_1 \ldots a_{2k}}$ at $k=0$ in the eq.(3). The first term in
the right hand side of eq.(15) corresponds to the well known one-loop
renormalization of the dilaton. The second one leads to the following
one-loop divergences of effective action in sector of the first massive level
background fields:
\begin{eqnarray}
   \Gamma_{div}^{(1)} =
    -  {\mu^\epsilon \over \epsilon}  \; \alpha' \! \int \! d^2 \sigma
     \sum_{i_1 = 1}^{18} {\cal O}_{i_1}^{(1)} \partial^2 B_{i_1}^{(1)},
\end{eqnarray}
where $\partial^2 = \eta^{\mu\nu} \partial_\mu \partial_\nu$.
Hence the one-loop renormalization of massive background fields in linear
approximation looks like that:
\begin{eqnarray}
\stackrel{\circ}{B}{}^{(1)}_{i_1}  =
       \mu^{-\epsilon}    \left(     B^{(1)}_{i_1} +
       {1\over{\epsilon}} \partial^2   B^{(1)}_{i_1} \right), \qquad
                        i_1 = 1,2,\ldots,18
\end{eqnarray}
Here the index $\circ$ denotes the bare fields. We see that the
renormalization has quite the universal structure.

To apply the dimensional regularization we formulate the theory on
$(2+\epsilon)$-dimensional world sheet. Then the trace of energy-momentum
tensor should have the general form:
\begin{eqnarray}
T \sim
\alpha' \sum_{i_1}^{18} T_{i_1}^{(1)} {\cal O}_{i_1}^{(1)}
+ {1\over 4} \; \epsilon \; \sqrt{g} g^{ab}
\partial_a x^\mu \partial_b x^\nu \delta_{\mu\nu} + \ldots,
\end{eqnarray}
where $T^{(1)}_{i_1}$ are functions of the background fields $B^{(1)}_{i_1}$
and their derivatives. Here dotes denote terms which do not contribute to the
quantum average values in massive background field sector. In accordance with
the eq.(18) the trace operator is a linear combination of two- and
four-dimensional operators. Hence to construct renormalized trace operator we
have to carry out the renormalization of corresponding composite operators and
express all bare background fields through renormalized ones taking into
account the eqs.(17).

The situation with  renormalization of dimension four composite operators
is the same as with renormalization of the effective action. A bare composite
operator is written  as a sum of infinite number of renormalized composite
operators of any dimension. However if we want to follow given massive level
contribution to the quantum trace operator it is sufficient to take into
account only finite number of renormalized composite operators.

Let us start with the dimension four composite operators.Consider,
 for example, the operator
$g^{ab} g^{cd} \partial_a x^\mu \partial_b x^\nu \partial_c x^\lambda
\partial_d x^\kappa f_{\!\mu\nu\lambda\kappa} (x)$,
where $f_{\!\mu\nu\lambda\kappa}$ is some combination of background fields
and their derivatives. We write $x^\mu = \bar x^\mu + \xi^\mu$ where $\bar
x^\mu$ are background coordinates and $\xi^\mu$ are quantum ones. In one-loop
approximation we have to consider terms of second order in $\xi^\mu$ only.
Then we should calculate average value and extract the divergent part. As a
result we will have in linear approximation:
\begin{eqnarray}
& & \left(
 g^{ab} g^{cd} \partial_a x^\mu     \partial_b x^\nu
\partial_c x^\lambda \partial_d x^\kappa
\stackrel{\circ}{f}\!{}_{\mu\nu\lambda\kappa}
\right)_0 =
\nonumber\\& &
= \left[
g^{ab} g^{cd} \partial_a x^\mu     \partial_b x^\nu
\partial_c x^\lambda \partial_d x^\kappa
\left(
\stackrel{\circ}{f}\!{}_{\mu\nu\lambda\kappa} (x)
-
{\mu^\epsilon \over \epsilon}
\partial^2 \! \stackrel{\circ}{f} \! {}_{\mu\nu\lambda\kappa} (x)
\right)
\right].
\end{eqnarray}
Here brackets denote renormalized composite operators. Now we express bare
background fields on right hand side of eq.(19) in terms of renormalized
ones:
$\stackrel{\circ}{f} = \mu^\epsilon ( f + {1 \over \epsilon} \partial^2 \! f)$
(see the eq.(17). As a result we obtain
\begin{eqnarray}
\left(
 g^{ab} g^{cd} \partial_a x^\mu     \partial_b x^\nu
               \partial_c x^\lambda \partial_d x^\kappa
               \stackrel{\circ}{f}{}_{\mu\nu\lambda\kappa}
\right)_0
   =
\left[
g^{ab} g^{cd} \partial_a x^\mu     \partial_b x^\nu
              \partial_c x^\lambda \partial_d x^\kappa
              f_{\mu\nu\lambda\kappa}
\right]
\end{eqnarray}
The analogous relations take place for any other dimension four composite
operators. In the linear approximation all dimension four composite operators
are finite.

Now let us consider the only dimension four operator
$g^{ab} \partial_a x^\mu \partial_b x^\nu \eta_{\mu\nu}$.
In this case straightforward calculations leads to the following result:
\begin{eqnarray}
& &
\left(
g^{ab} \partial_a x^\mu \partial_b x^\nu \eta_{\mu\nu}
\right)_0
  =
\left[
g^{ab} \partial_a x^\mu \partial_b x^\nu \eta_{\mu\nu}
\right]
+ {{\mu^\epsilon} \over {\epsilon}} A,
\nonumber \\& &
A =
\Bigl[
      P_\mu{}^\mu - {1\over{2}} D_a P^a{}_\mu{}^\mu
      + {1\over{2}} D_a D_b  P^{ab}{}_\mu{}^\mu
      - {1\over{4}} g_{ab} D^2 P^{ab}{}_\mu{}^\mu
      + {1\over{2}} D^a
      \left(
              R^{(2)} P_{1\;a}{}^\mu{}_\mu
      \right)
\nonumber\\{}& &
\qquad\qquad
     + {1\over{6}} D_a
       \left(
              R^{(2)} \varepsilon^{ab} P_{2\;b}{}^\mu{}_\mu
       \right)
     + {1\over{2}} D^a
       \left(
              \tilde P_\mu{}^\mu D_a R^{(2)}
       \right)
\Bigr]
\end{eqnarray}
where
\begin{eqnarray}
& &     P^{abc}{}_{\mu\nu} =
        \left(
              g^{ab} g^{cd} + g^{ac} g^{bd} + g^{ad} g^{bc}
        \right)
            P_{1\;d\;\mu\nu}
        + g^{(ab} \varepsilon^{c)d} P_{2\;d\;\mu\nu},
\nonumber \\ & &
        P^{abcd}{}_{\mu\nu} =
\left(  g^{ab} g^{cd} + g^{ac} g^{bd} + g^{ad} g^{bc} \right)
        \tilde P_{\mu\nu}
\end{eqnarray}
The functions $P_{\mu\nu}$, $P^{a}{}_{\mu\nu}$, $P^{ab}{}_{\mu\nu}$,
$P^{abc}{}_{\mu\nu}$, $P^{abcd}{}_{\mu\nu}$ are given by the eqs.(3,4) as the
coefficients of the differential operator $\cal H$ at $n=1$.

\vspace{0.5cm}

{\bf 5. Background Fields Equation of Motion.}
As a result the renormalized trace of the energy-momentum tensor has the form
\begin{eqnarray}
[ T ] \sim
\alpha' \left. \sum_{i_1} \right.^\prime T_{i_1}^{(1)}
\left[ {\cal O}_{i_1}^{(1)} \right]
+ A  \equiv
\left. \sum_{i_1} \right.^\prime E_{i_1}^{(1)}
 \left[ {\cal O}_{i_1}^{(1)} \right],
\end{eqnarray}
where the prime indicates that there are no terms proportional to $D^2 x^\mu$
in the summation. The requirement of the quantum Weyl invariance is $[T]=0$.
It means that all functions $E^{(1)}_{i_1}$ should be equal to zero. Thus we
obtain the equation of motion for the first massive level background fields
$E^{(1)}_{i_1} = 0$. After straightforward calculations we are able to write
these equations in the explicit form:
\begin{eqnarray}
& &\partial^2  F^{(1)}_{\mu\nu\lambda\kappa}
- 4 \partial_\mu \partial^\alpha    F^{(1)}_{\alpha\nu\lambda\kappa}
+ 2 \partial_\mu \partial_\nu    F^{(1)}_{\lambda\kappa\alpha}{}^\alpha
+ 2 \partial_\mu \partial_\nu    F^{(1)}_{\lambda\alpha}{}^\alpha{}_\kappa
+ \partial_\mu \partial_\lambda \partial^\alpha   T^{(2)}_{\alpha\nu\kappa}
\nonumber \\ & & {}\qquad
- {1 \over 2} \partial_\mu \partial_\nu \partial^\alpha
                                    T^{(2)}_{\alpha\lambda\kappa}
- \partial_\mu \partial_\nu \partial_\lambda   T^{(1)}_\alpha{}^\alpha_\kappa
+ {1 \over 2} \partial_\mu \partial_\nu \partial_\lambda \partial_\kappa
                                    M^{(2)}_\alpha{}^\alpha
\nonumber \\ & &\qquad {}
- m^2  \left(
  F^{(1)}_{\mu\nu\lambda\kappa}
+ {1 \over 2} \partial_\mu T^{(2)}_{\nu\lambda\kappa}
- {1 \over 2} \partial_\mu T^{(2)}_{\lambda\nu\kappa}
- {1 \over 2} \partial_\mu T^{(2)}_{\lambda\kappa\nu}
- \partial_\mu \partial_\nu W^{(1)}_{\lambda\kappa}
\right. \nonumber \\ & &  \left. {}
\qquad
+ \partial_\mu \partial_\nu \partial_\lambda Y^{(1)}_\kappa
- \partial_\mu \partial_\nu \partial_\lambda \partial_\kappa Q
\right)
= 0 \nonumber \\
& &\partial^2 F^{(2)}_{\mu\nu\lambda\kappa}
+ 6 \partial_\lambda \partial^\alpha F^{(2)}_{\mu\nu\kappa\alpha}
- 4 \partial_\lambda \partial^\alpha F^{(2)}_{\kappa\mu\nu\alpha}
+ 4 \partial_\lambda \partial_\mu F^{(2)}_{\nu\alpha}{}^\alpha{}_\kappa
\nonumber \\ & & {}\qquad
- 4 \partial_\lambda \partial_\mu F^{(2)}_{\kappa\alpha}{}^\alpha_\nu
+ \partial_\nu \partial_\mu \partial^\alpha T^{(3)}_{\alpha(\nu\kappa)}
- \partial_\lambda \partial_\mu \partial_\nu T^{(3)\alpha}{}_{(\alpha\kappa)}
\nonumber \\ & & {}\qquad
- m^2  \left(
  F^{(2)}_{\mu\nu\lambda\kappa}
+ {1 \over 2} \partial_\lambda T^{(3)}_{\kappa(\mu\nu)}
- \partial_\lambda T^{(3)}_{\mu(\nu\kappa)}
- \partial_\mu \partial_\nu W^{(2)}_{\lambda\kappa}
- \partial_\mu \partial_\nu \partial_\lambda Y^{(2)}_\kappa
\right)
= 0 \nonumber \\
& &\partial^2 T^{(2)}_{\mu\nu\lambda}
- 2 \partial_\lambda \partial^\alpha T^{(2)}_{\mu\nu\alpha}
+ \partial_\mu \partial^\alpha T^{(2)}_{\alpha\nu\lambda}
- 2 \partial_\mu \partial_\nu T^{(2)}_\alpha{}^\alpha{}_\lambda
- 2 \partial_\nu \partial_\lambda T^{(2)}_\alpha{}^\alpha{}_\mu
\nonumber \\ & & {}\qquad
- 8 \partial^\alpha F^{(1)}_{\alpha\nu\mu\lambda}
+ 8 \partial_\lambda F^{(1)}_{\mu\nu\alpha}{}^\alpha
+ 8 \partial_\lambda F^{(1)}_{\mu\alpha}{}^\alpha{}_\nu
+ 2 \partial_\lambda \partial_\nu \partial^\alpha M^{(2)}_{\alpha\mu}
+ 2 \partial_\mu \partial_\nu \partial_\lambda M^{(2)}{}_\alpha{}^\alpha
\nonumber \\ & & {}\qquad
- m^2  \left(
- 2 \partial_\lambda M^{(2)}_{\mu\nu}
- 4 \partial_\lambda W^{(1)}_{\mu\nu}
+ 2 \partial_\lambda \partial_\mu Y^{(1)}_\nu
+ 2 \partial_\lambda \partial_\nu Y^{(1)}_\mu
- 4 \partial_\mu \partial_\nu \partial_\lambda Q
\right)
= 0 \nonumber \\
& &\partial^2 T^{(3)}_{\mu(\nu\lambda)}
+ \partial_\mu \partial^\alpha T^{(3)}_{\alpha(\nu\lambda)}
- \partial_\lambda \partial^\alpha T^{(3)}_{\mu(\nu\alpha)}
- \partial_\nu \partial^\alpha T^{(3)}_{\mu(\alpha\lambda)}
- \partial_\mu \partial_\lambda T^{(3)}{}^\alpha_{(\alpha\nu)}
\nonumber \\ & & {}\qquad
- \partial_\mu \partial_\nu T^{(3)}{}^\alpha{}_{(\alpha\lambda)}
+ 2 \partial_\nu \partial_\lambda T^{(3)}{}^\alpha_{(\alpha\mu)}
+ 4 \partial^\alpha F^{(2)}_{\mu\nu\lambda\alpha}
+ 4 \partial^\alpha F^{(2)}_{\mu\lambda\nu\alpha}
\nonumber \\ & & {}\qquad
- 4 \partial^\alpha F^{(2)}_{\nu\lambda\mu\alpha}
+ 8 \partial_\nu F^{(2)}_{\mu]\alpha}{}^\alpha{}_{[\lambda}
+ 8 \partial_\lambda F^{(2)}_{\mu]\alpha}{}^\alpha{}_{[\nu}
- 2 \partial_\nu \partial_\lambda \partial^\alpha M^{(3)}_{\alpha\mu}
\nonumber \\ & & {}\qquad
- m^2  \left(
- \partial_\lambda M^{(3)}_{\mu\nu}
- \partial_\nu M^{(3)}_{\mu\lambda}
- 2 \partial_\nu W^{(2)}_{\mu\lambda}
- 2 \partial_\lambda W^{(2)}_{\mu\nu}
\right. \nonumber \\ & & \left. {}
\qquad
- \partial_\mu \partial_\nu Y^{(2)}_\lambda
- \partial_\mu \partial_\lambda Y^{(2)}_\nu
+ 2 \partial_\lambda \partial_\nu Y^{(2)}_\mu
\right)
= 0 \nonumber \\
& &\partial^2 M^{(2)}_{\mu\nu}
+ 2 \partial_\mu \partial^\alpha M^{(2)}_{\alpha\nu}
+ \partial_\mu \partial_\nu M^{(2)}{}_\alpha{}^\alpha
+ 4 F^{(1)}_{\mu\nu\alpha}{}^\alpha
+ 4 F^{(1)}_{\mu\alpha}{}^\alpha{}_\nu
\nonumber \\ & & {}\qquad
- 2 \partial^\alpha T^{(2)}_{\mu\nu\alpha}
- 2 \partial_\mu T^{(2)}{}_\alpha{}^\alpha{}_\nu
\nonumber \\ & & {}\qquad
+ m^2 \left(
 M^{(2)}_{\mu\nu}
+ 2 W^{(1)}_{\mu\nu}
- 2 \partial_\mu Y^{(1)}_\nu
+ \partial_\mu \partial_\nu Q
\right)
= 0 \nonumber \\
& &\partial^2 M^{(3)}_{\mu\nu}
- 2 \partial_\nu \partial^\alpha M^{(3)}_{\alpha\mu}
+ 8 F^{(2)}_{\mu]\alpha}{}^\alpha{}_{[\nu}
- 2 \partial^\alpha T^{(3)}_{\mu(\nu\alpha)}
+ 2 \partial_\nu T^{(3)}{}^\alpha{}_{(\alpha\mu)}
\nonumber \\ & & {}\qquad
+ m^2 \left(
M^{(3)}_{\mu\nu}
+ 2 W^{(2)}_{\mu\nu}
+ 2 \partial_\mu Y^{(2)}_\nu
\right)
= 0 \nonumber \\
& &\partial^2 W^{(1)}_{\mu\nu}
- 2 \partial_\nu \partial^\alpha W^{(1)}_{\alpha\mu}
+ \partial_\mu \partial_\nu W^{(1)}{}_\alpha{}^\alpha
- 2 F^{(1)}_{\mu\nu\alpha}{}^\alpha
- 2 F^{(1)}_{\mu\alpha}{}^\alpha{}_\nu
\nonumber \\ & & {}\qquad
+ \partial^\alpha T^{(2)}_{(\mu\nu)\alpha}
+ \partial_\mu T^{(2)}{}_\alpha{}^\alpha{}_\nu
- 2 \partial_\mu \partial^\alpha M^{(2)}_{\alpha\nu}
\nonumber \\ & & {}\qquad
- m^2 \left(
M^{(2)}_{\mu\nu}
+ 2 W^{(1)}_{\mu\nu}
- 2 \partial_\mu Y^{(1)}_\nu
+ 2 \partial_\mu \partial_\nu Q
- 2 \partial_\mu \partial_\nu C
\right)
= 0 \nonumber \\
& &\partial^2 W^{(2)}_{\mu\nu}
- 2 \partial_\mu \partial^\alpha W^{(2)}_{\alpha\nu}
- 2 F^{(2)}_{\mu\alpha}{}^\alpha{}_\nu
+ 2 F^{(2)}_{\nu\alpha}{}^\alpha{}_\mu
+ \partial^\alpha T^{(3)}_{\mu(\alpha\nu)}
\nonumber \\ & & {}\qquad
- \partial_\nu T^{(3)}{}^\alpha{}_{(\alpha\mu)}
- 2 \partial_\mu \partial^\alpha M^{(3)}_{\alpha\nu}
- m^2 \left(
M^{(3)}_{\mu\nu}
+ 2 W^{(2)}_{\mu\nu}
+ 2 \partial_\mu Y^{(2)}_\nu
\right)
= 0 \nonumber \\
& &\partial^2 Y^{(1)}_\mu
- \partial_\mu \partial^\alpha Y^{(1)}_\alpha
- \partial^\alpha M^{(2)}_{\alpha\mu}
+ \partial_\mu M^{(2)}{}_\alpha{}^\alpha
- 2 \partial^\alpha W^{(1)}_{\alpha\mu}
+ 2 \partial_\mu W^{(1)}{}_\alpha{}^\alpha
+ m^2 4 \partial_\mu C
= 0 \nonumber \\
& &\partial^2 Y^{(2)}_\mu
- \partial_\mu \partial^\alpha Y^{(2)}_\alpha
+ \partial^\alpha M^{(3)}_{\alpha\mu}
+ 2 \partial^\alpha W^{(2)}_{\alpha\mu}
= 0 \nonumber \\
& &\partial^2 Q
+ {1 \over 2} M^{(2)}{}_\alpha^\alpha
+ W^{(2)}{}_\alpha^\alpha
- \partial^\alpha Y^{(1)}_\alpha
+ m^2 2 C
= 0 \nonumber \\
& &\partial^2 C
- m^2 C
= 0
\end{eqnarray}
Here $m$ is the mass of the first massive level of closed string. It can be
shown that these equations are invariant under the gauge transformations
(12). Hence we have a possibility to fix the invariance by setting all the
Stueckelberg fields equal to zero. Then the eqs.(24) will have the form
\begin{eqnarray}
& &\left( \partial^2 - m^2 \right) F^{(1)}_{\mu\nu\lambda\kappa} = 0,
\qquad
\left( \partial^2 - m^2 \right) F^{(2)}_{\mu\nu\lambda\kappa} = 0,
\nonumber \\ & &
\left( \partial^2 - m^2 \right) W^{(1)}_{\mu\nu} = 0,
\qquad
\left( \partial^2 - m^2 \right) W^{(2)}_{\mu\nu} = 0,
\qquad
\left( \partial^2 - m^2 \right) C = 0,
\\
& &\partial^\alpha F^{(1)}_{\alpha\mu\nu\lambda} = 0,
\qquad
\partial^\alpha F^{(2)}_{\alpha\mu\nu\lambda} = 0,
\qquad
\partial^\alpha W^{(1)}_{\alpha\mu} = 0,
\qquad
\partial^\alpha W^{(2)}_{\alpha\mu} = 0,
\\
& &2 F^{(1)}_{\mu\nu\alpha}{}^\alpha +
2 F^{(1)}_{\mu\alpha}{}^\alpha{}_\nu + m^2 W^{(1)}_{\mu\nu} = 0,
\qquad
4 F^{(2)}_{\mu\alpha}{}^\alpha{}_\nu -
4 F^{(2)}_{\nu\alpha}{}^\alpha{}_\mu + 2 m^2 W^{(2)}_{\mu\nu} = 0,
\nonumber \\& &
W^{(1)}{}_\alpha{}^\alpha + 2 m^2 C = 0.
\end{eqnarray}
The equations (25-27) are final.

Let us consider these equations in details. The eqs.(25) define the standart
on-shell conditions. The eqs.(26) are the transversality conditions. To
clarify the content of these eqs.(25-27) we introduce
\begin{eqnarray}
& & F_{\mu\nu\,\lambda\kappa} =
2 F^{(1)}_{\mu\lambda\nu\kappa} + 2 F^{(1)}_{\mu\kappa\nu\lambda}
- 2 F^{(2)}_{\mu\lambda\nu\kappa} - 2 F^{(2)}_{\nu\kappa\mu\lambda},
\nonumber \\ & &
 W_{\mu\,\nu} = - W^{(1)}_{\mu\nu} + W^{(2)}_{\mu\nu},
\nonumber \\ & &
F_{\mu\nu\,\lambda\kappa} = F_{(\mu\nu)\,(\lambda\kappa)}
\end{eqnarray}
Then the eqs.(25-27) are rewritten as
\begin{eqnarray}
& &\left( \partial^2 - m^2 \right)  F_{\mu\nu\,\lambda\kappa} = 0
\\
& &\partial^\alpha F_{\alpha\mu\,\nu\lambda} = 0,
\qquad
\partial^\alpha F_{\mu\nu\,\lambda\alpha} = 0
\\
& & W_{\mu\,\nu} =  - {2 \over m^2} F_{\mu\alpha\,}{}^\alpha{}_\nu ,
\qquad
 C = - {1 \over 2m^2} W_{\alpha\,}{}^\alpha
\end{eqnarray}
The eqs.(31) mean that the fields $W_{\mu\nu}$ and $C$ are not independent
ones. They present simply the denotion for the traces
$F_{\mu\alpha}{}^\alpha{}_\nu$ and $F_{\mu\alpha}{}^{\alpha\mu}$ respectively.
As a result the linear background field equations of motion have the form
(29,30).

It is well known that the first massive level of closed bosonic string
corresponds to a tensor field of fourth rank $f_{\mu\nu\alpha\beta} =
f_{(\mu\nu)(\alpha\beta)}$ satisfying the
on-shell equations (29) and also the conditions of transversality (30) and
of tracelessness inside two pairs of indices.
Thus the theory under consideration will correspond to the
first massive string excitations if we impose the additional requirements
$F_{\mu\nu\alpha}{}^\alpha = 0$, $F_{\alpha}{}^\alpha{}_{\mu\nu} = 0$ on the
background fields in the lagrangian (9)\footnote{The problem of tracelessness
condition has been established in approach of exact renormalization group for
open string in \cite{8}. It has been noticed in \cite{10} that to obtain this
condition for open string one should consider two-loop approximation.}.

\vspace{0.5cm}

{\bf 6. Conclusion.}
In this paper we have presented the generalization of $\sigma$-model approach
allowing to describe the interaction of both massless and massive string
modes. The model including the background fields corresponding to the first
massive level has been considered in details.

We have shown that the renormalization of the theory with finite number of
massive level contributions is realized independently on higher massive
levels background fields. It allows to investigate the proposed model by
standart methods of the renormalization theory and to derive the renormalized
trace of the energy-momentum tensor. Then the requirement of quantum Weyl
invariance leads to equation of motion for all massless and massive
background fields. This programme has been realized in one-loop weak field
approximation and the result corresponds to the first massive level spectrum
of closed bosonic string under some restrictions on the background fields. We
believe that to get these restrictions in framework of our approach it is
sufficient to find two-loop weak field contribution to the quantum trace.

It is obvious that the proposed approach can as well be realized beyond
the weak field approximation, when interactions of massive modes with each
other and with massless modes are taken into account. Since the background
fields corresponding to massive string levels include fields of higher
spins we can expect that this approach opens the way to construct the theory
of interacting massive higher spins coupled to gravity.

\vspace{0.5cm}
{\bf Acknowledgements.}
I.L.B. is very grateful to Hugh Osborn for a warm hospitality at DAMTP,
University of Cambridge and fruitful discussions. He is thankful to
A.A.Tseytlin for stimulating detailed discussions and critical comments and
to C.Hull, P.Townsend, P.West for useful conversations.

\end{document}